\documentclass{aastex6}

\usepackage{savesym}          %  There is a conflict of amsmath and texfonts
\savesymbol{iint}             %  Both define i...int which in my installation 
\savesymbol{iiint}            %  causes an error. These additional lines rename
\savesymbol{iiiint}           %  the amsmath definitions to:
\savesymbol{idotsint}         %  
\usepackage{amsmath}
\restoresymbol{AMS}{iint}     %  -> AMSiint        \  these amsmath integral
\restoresymbol{AMS}{iiint}    %  -> AMSiiint        \  symbols are inferior 
\restoresymbol{AMS}{iiiint}   %  -> AMSiiiint       /  to those from txfonts 
\restoresymbol{AMS}{idotsint} %  -> AMSidotsint    /  We should not use them
\newcommand{\regpar}{\xi}
\usepackage{natbib}
\usepackage{color}
\begin{document}
\title{Nonlinear force-free coronal magnetic stereoscopy}
\date{Draft version December 20, 2016.}
\author{Iulia Chifu,$^{1,2}$
Thomas Wiegelmann$^{1}$,
Bernd Inhester$^{1}$}
\affil{
$^{1}$ Max-Planck-Institut f\"ur Sonnensystemforschung, Justus-von-Liebig-Weg 3,
37077 G\"ottingen, Germany;\\ chifu@mps.mpg.de\\
$^{2}$ Astronomical Institute of Romanian Academy, Cutitul de Argint 5,
Bucharest, Romania \\}

\begin{abstract}
Getting insights into the 3D structure of the solar coronal magnetic
field have been done in the past by two completely different approaches:
(1.) Nonlinear force-free field (NLFFF) extrapolations, which use photospheric
vector magnetograms as boundary condition. (2.) Stereoscopy of coronal magnetic loops 
observed in EUV coronal images from different vantage points.
Both approaches have their strength and weaknesses. Extrapolation methods are
sensitive to noise and inconsistencies in the boundary data and the accuracy
of stereoscopy is affected by the ability of identifying the same
structure in different images and by the separation angle between 
the view directions. As a consequence, for the same observational data, 
the computed 3D coronal magnetic field with the two methods 
do not necessarily coincide. In an earlier work (Paper I) we
extended our NLFFF optimization code by the inclusion of stereoscopic
constrains. The method was successfully tested with synthetic data and
within this work we apply the newly developed code to a combined data-set
from SDO/HMI, SDO/AIA and the two STEREO spacecraft. The extended method
(called S-NLFFF) contains an additional term that monitors and minimizes
the angle between the local magnetic field direction and the orientation of
the 3D coronal loops reconstructed by stereoscopy.
We find that prescribing the shape of the 3D stereoscopically reconstructed loops
the S-NLFFF method leads to a much better agreement between the modeled field and the stereoscopically 
reconstructed loops. We also find an appreciable decrease by a factor of two in the angle between the current and the magnetic field 
which indicates the improved quality of the force-free solution obtained by S-NLFFF.
\end{abstract}
\keywords{Sun: corona, Sun: magnetic fields, methods: numerical}

\section{Introduction} \label{sec:intro}
Knowledge of the 3D structure of the solar coronal magnetic
field is essential to understand basically all physical
processes in the corona. The reason is that the magnetic
field clearly dominates and structures the corona, because
the plasma $\beta$ (ratio of plasma and magnetic pressure) is very small. Unfortunately direct measurements of the coronal magnetic field are not routinely available and two
distinct methods have been developed to reconstruct the coronal magnetic field: 1.) extrapolations of photospheric vector fields
into the corona under the force-free assumption
\citep[see][for a review]{WiegelmannSakurai2012} and
2.) Stereoscopy of coronal images
\citep[see][for a review]{2011LRSP....8....5A}.
Both methods are not perfect if applied to observational
data. Photospheric vector magnetograms contain noise and
are not necessarily force-free consistent because of the mixed plasma
$\beta$ in the lower solar atmosphere \citep{1990MmSAI..61..457G}. For a stereoscopic
reconstruction from different vantage points one first has to extract
loop-like structures from EUV-images, identify the same loop
in both images (association problem) and finally perform the 3D stereoscopy
(large error at loop-top for East-West loops). Consequently the output
of NLFFF and stereoscopy can be different \citep[see][for a comparison of NLFFF-models and stereoscopy]{2009ApJ...696.1780D}.

It is therefore natural to combine photospheric measurements and stereoscopy
to obtain coronal magnetic field measurements which comply
with both data sets. Several such attempts have been made,
whereas the methods developed so far use the photospheric
line-of-sight field, rather than the full vector field, as boundary
condition.
 First attempts have been made about one and a half decade
ago by \cite{2002SoPh..208..233W} using linear force-free fields
with SOHO/MDI magnetograms as boundary conditions. In this approach
the linear force-free parameter $\alpha$ was computed by comparing
the resulting fields with 3D-loops from dynamic stereoscopy
\citep[see][]{1999ApJ...515..842A}. That time, well before the launch of STEREO,
images from different vantage points have been observed using the rotation of the Sun, and it was therefore necessary to limit the method to almost
stationary structures. The method was later extended by
\cite{2003SoPh..218...29C} to compute the linear force-free $\alpha$
also directly from coronal images from one viewpoint only.
In subsequent works, still within the limitations of linear force-free
models, projections of the magnetic field loops have been
used to solve the stereoscopic association and ambiguity problem.
The method was dubbed {\it magnetic stereoscopy} 
\citep[see][for details]{2006SoPh..236...25W,2007ApJ...671L.205F}

Linear force-free fields have their limitation
\citep[see, e.g.,][]{2008JGRA..113.3S02W} and in particular the best fit value of
$\alpha$ for different loops within one active region are different
and $\alpha$ can even change it's sign. \cite{2012ApJ...756..124A}
incorporated a forward fitting method, which uses analytic expressions
and  different values of $\alpha$ along different loops, thereby
approximating a nonlinear force-free field. The method was refined in
\cite{2013SoPh..287..323A,2013ApJ...763..115A} and subsequent code versions allow
using 2D-loop projections rather
than 3D-stereo-loops. The method was intensively tested, compared
with extrapolations from vector magnetograms and further refined
in a number of subsequent paper,
\citep[e.g.][]{2013SoPh..287..345A,2013SoPh..287..369A,2014ApJ...785...34A,
2016ApJS..224...25A}. It was dubbed
{\it Vertical-Current Approximation Nonlinear Force-Free Field
(VCA-NLFFF) code}.

While VCA-NLFFF avoids several problems of
magnetic field extrapolations from photospheric vector magnetograms, e.g.
the assumption that the boundary data are force-free consistent is not
necessary, the method uses only the line-of-sight photospheric magnetic
field and not the full vector field.

\cite{2012ApJ...756..153M, 2014ApJ...783..102M} proposed a NLFF field extrapolation method, called Quasi-Grad-Rubin, which uses the line-of-sight component of the surface magnetic field and the 2D shapes of the coronal loops from a single image as constraints for their extrapolation. They tested the method with a semi-analytic solution and also applied it on observational data.

Within this work, we propose a new
method which we call {\it Stereoscopic Nonlinear force-free field code
(S-NLFFF)}. The method uses both photospheric vector magnetograms (here from
SDO/HMI) and stereoscopic reconstructed 3D-loops as input.
Necessarily providing all these conditions over-imposes the boundary
condition and one cannot find a solution which strictly fulfills
constraints which probably contradicts each other.
The advantage of our new method is that the different constraints
(force-freeness, photospheric magnetic field vector, 3D-stereo-loops)
are all considered as terms of one functional, each weighted
with certain Lagrangian multipliers. These free parameters allow
to specify measurement errors (both in the photospheric field as well
as in the  prescribed 3D-loops) and the code iterates for an optimal
solution in the sense that deviation from the boundary conditions are allowed
in regions with a substantial measurement error (photospheric field vector)
and reconstruction error (stereo-loops). The method was described and
tested with synthetic data in \cite{ChifuEtal2015} (Paper-I).

The paper is outlined as follows: in section 2 we make a short description of the methods used for the reconstruction of the 3D coronal loops and of the 3D magnetic field, in section 3 we present the data used for the reconstructions, in section 4 we show the 3D reconstruction, in section 5 we present the results and in section 6 we discuss the results.

\section{Methods}\label{sec:methods}

\subsection{Multiview B-spline Stereoscopic Reconstruction (MBSR)}\label{subsec:MBSR} 

The 3D shape of solar loop-like structures (e.g. coronal loops, prominences, leading edge of coronal mass ejections) can be performed using stereoscopic reconstruction. Two-view directions are sufficient for a 3D reconstruction from an ideal data set. The use of more views brings more accuracy to the reconstruction if the data are noisy. The main steps in the stereoscopic reconstruction are: the identification of the object to be reconstructed in all of the available views; matching the object by tie-pointing; the reconstruction \citep{Inhester2006}. Usually, as a final step the stereoscopically reconstructed points from the loop-like structure often needs to be smoothed by fitting a polynomial or a spline curve \citep{Chifu2016}. 

The main idea of the MBSR method is the reconstruction in one go of an entire loop-like structure. Instead of calculating pairwise reconstructions from multiple views which in the end needs to be averaged, our code is able to reconstruct tie-pointed curves from two or more views directly. The tie-points do not have to be related by a common epipolar coordinate and therefore be used directly in more than 2 views. It is designed to yield a unique 3D B-spline as approximation to the reconstructed loop curve, the projections of which optimally matches all tie-points in all images. The local error depends only on the projected distances of the tie-points position to the final spline curve \citep{Chifu2016}.

\subsection{Stereoscopic-Nonlinear Force-Free Field extrapolation (S-NLFFF) }\label{subsec:SNLFFF}
The modeling of the magnetic field in the solar corona is possible under certain assumptions.
The plasma $\beta$ model by \cite{Gary2001} shows that in the corona the magnetic pressure dominates
over the plasma pressure  and gravity effects and the kinematic ram pressure of plasma flows are small \citep{ WiegelmannSakurai2012}, too.
In this approach, called the force-free field assumption, the Lorentz-force vanishes
and has to fulfill the non-linear equation ($ \mathbf{j} \times \mathbf{B} = 0$) 
together with the solenoidal condition ($ \nabla \cdot \mathbf{B} = 0 $).
 
To model the coronal magnetic field using nonlinear force-free field extrapolations,
one needs surface observations of all three components of the magnetic field as boundary condition.
We solve the force-free equations with the help of an optimization approach, which has originally been proposed by \cite{WheatlandEtal2000}  and extended by \cite{Wiegelmann04,Wiegelmann10}. Recently, the NLFFF optimization method was extended by constraining the magnetic field to be aligned to the 3D coronal loops stereoscopically reconstructed from EUVI images \citep{ChifuEtal2015}.

The essential approach of the extended S-NLFFF method is to minimize a scalar cost function ($ \mathrm{L_{tot}}$) which consists of a number of terms quantifying constraints the final solution should satisfy. The terms of the functional are
\begin{gather}
  \text{L}_\textit{1}=\int_V w_f \frac{|(\nabla \times \mathbf{B}) \times
    \mathbf{B}|^2}{B^2} \;d^3r,
  \label{L1}\\
  \text{L}_\textit{2}=\int_V w_f |\nabla \cdot \mathbf{B}|^2 \;dr^3,
  \label{L2}\\
  \text{L}_\textit{3}=\int_S ( \mathbf{B} - \mathbf{B}_{obs} )
           \cdot \mathrm{diag(\sigma^{-2}_\alpha)}
           \cdot (\mathbf{B}-\mathbf{B}_{obs}) \;d^2r,
  \label{L3}\\
   \text{L}_\textit{4} = \sum_i \int_{\mathbf{c}_i} \frac{1}{\sigma^2_c}
  {|\mathbf{B}\times\mathbf{t}_i|^2}\;ds,
  \label{L4} \\
  \text{where}\quad
  \mathbf{t}_i=\frac{d\mathbf{c}_i}{ds}.
\end{gather}
The function to be minimized is 
\begin{equation}
  \mathrm{L_{tot}}=\sum_n \regpar_n L_n,
  \label{Ltot}
\end{equation}
where $\xi_i$ are regularization weights. Our experience from \cite{ChifuEtal2015} suggests $\xi_i = 1$ as an acceptable choice for the weights. 

The computational box has an inner physical domain surrounded by a buffer zone on the top and lateral boundaries. The force-free and divergence-free conditions are satisfied if the first two terms (Eq. \ref{L1} and  \ref{L2}) are minimized to zero. $w_f$ is a boundary weight function which is set to unity in the physical domain and it decreases monotonically to zero towards the outer buffer zone \cite[see][for more details]{Wiegelmann04}. The third term (Eq. \ref{L3}) minimizes the differences between the observed and modeled magnetic field at the bottom boundary, while the fourth term (Eq. \ref{L4}) minimizes the angles between the modeled magnetic field and the tangents of the stereoscopically reconstructed loops. In Eq. \ref{L3}, $\sigma_q(\mathbf{r})$ are estimated measurement errors for the three field components $\textit{q}=x,y,z$ on $S$ \cite[see][for more details]{TadesseEtal2011}. In Eq. \ref{L4}, $\sigma_{c_i}(s)$ is a relative measure of the estimated error of the tangent direction $\mathbf{t}_i(s)$ along the loop $i$. A detailed description of the NLFFF optimization method (the L$_\textit{1}$, L$_\textit{2}$, L$_\textit{3}$ terms) can be found in \cite{WheatlandEtal2000,Wiegelmann04,Wiegelmann10} and about S-NLFFF method (the L$_\textit{4}$ term) can be found in \cite{ChifuEtal2015}.

\section{Observational data} \label{sec:obs}

\begin{figure}[h]
		\centerline{\hspace*{0.001\textwidth}
			    }   
 \vspace{0.01\textwidth} 
 \centerline{\hspace*{0.001\textwidth}
		     \includegraphics[ width=0.32\textwidth, trim = 40 140 90 220, clip]{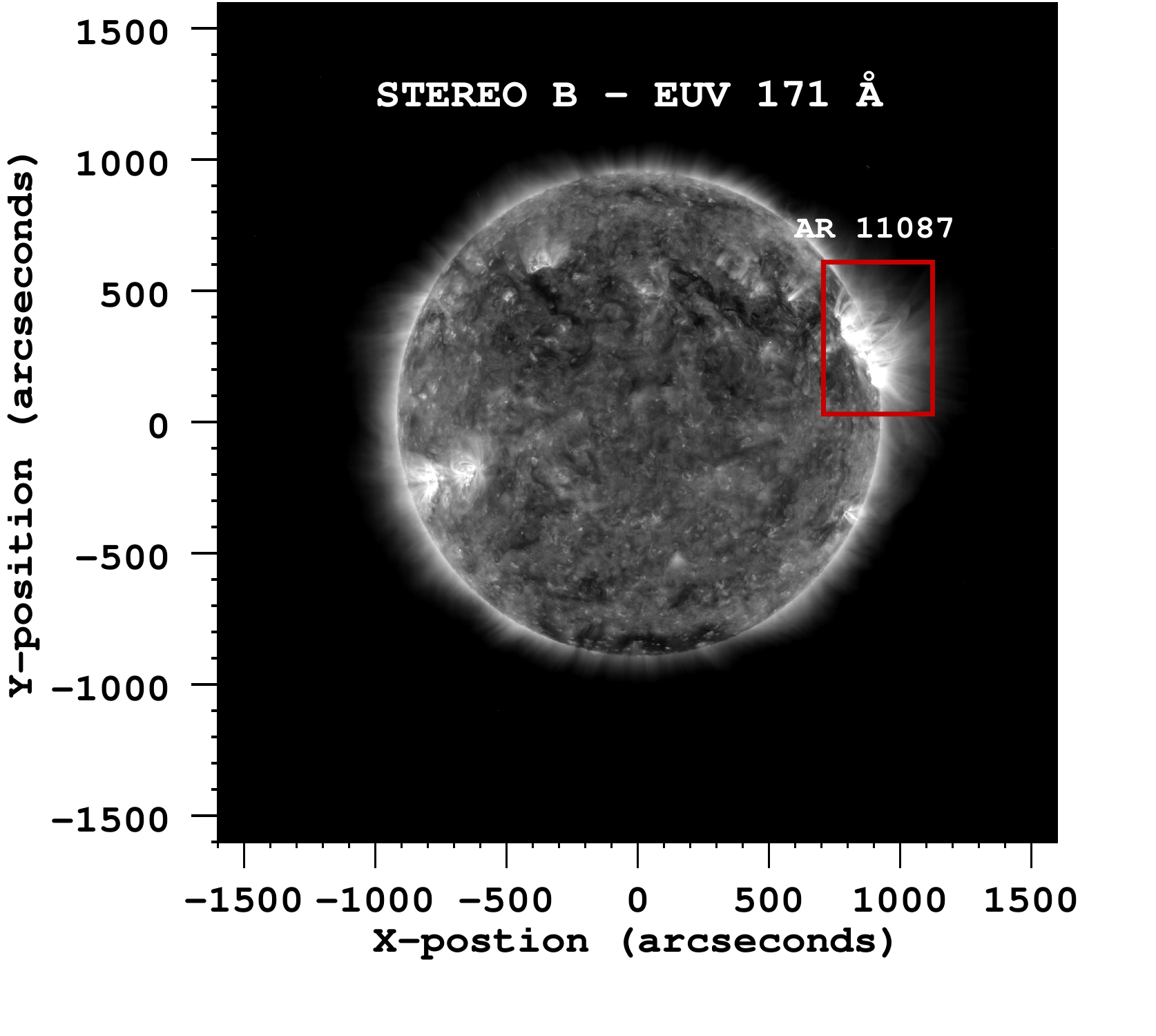} %trim=left bottom right top,clip
		       \includegraphics[width=0.32\textwidth, trim = 40 140 90 220, clip]{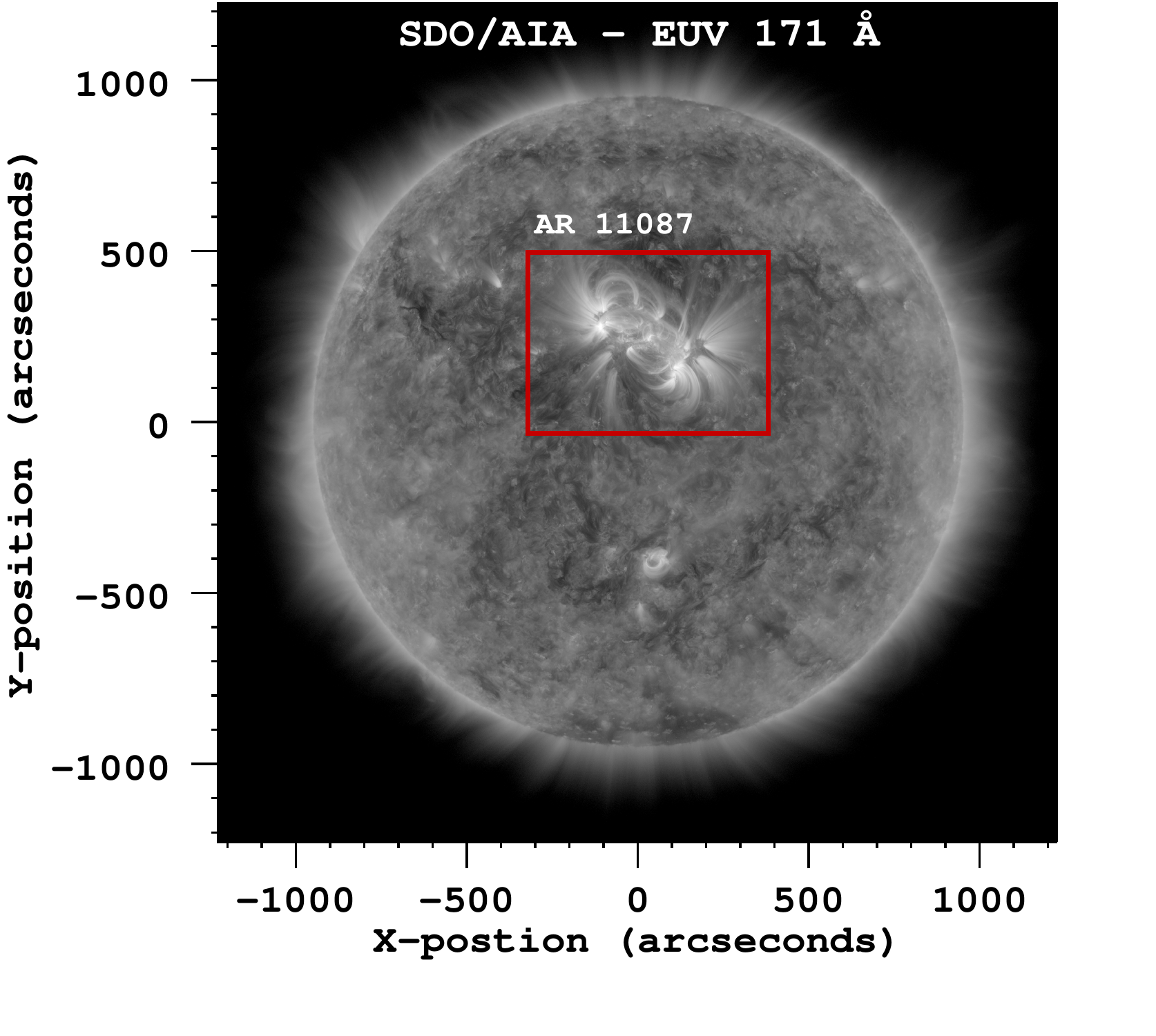}
		      \includegraphics[width=0.32\textwidth, trim = 40 140 90 220, clip]{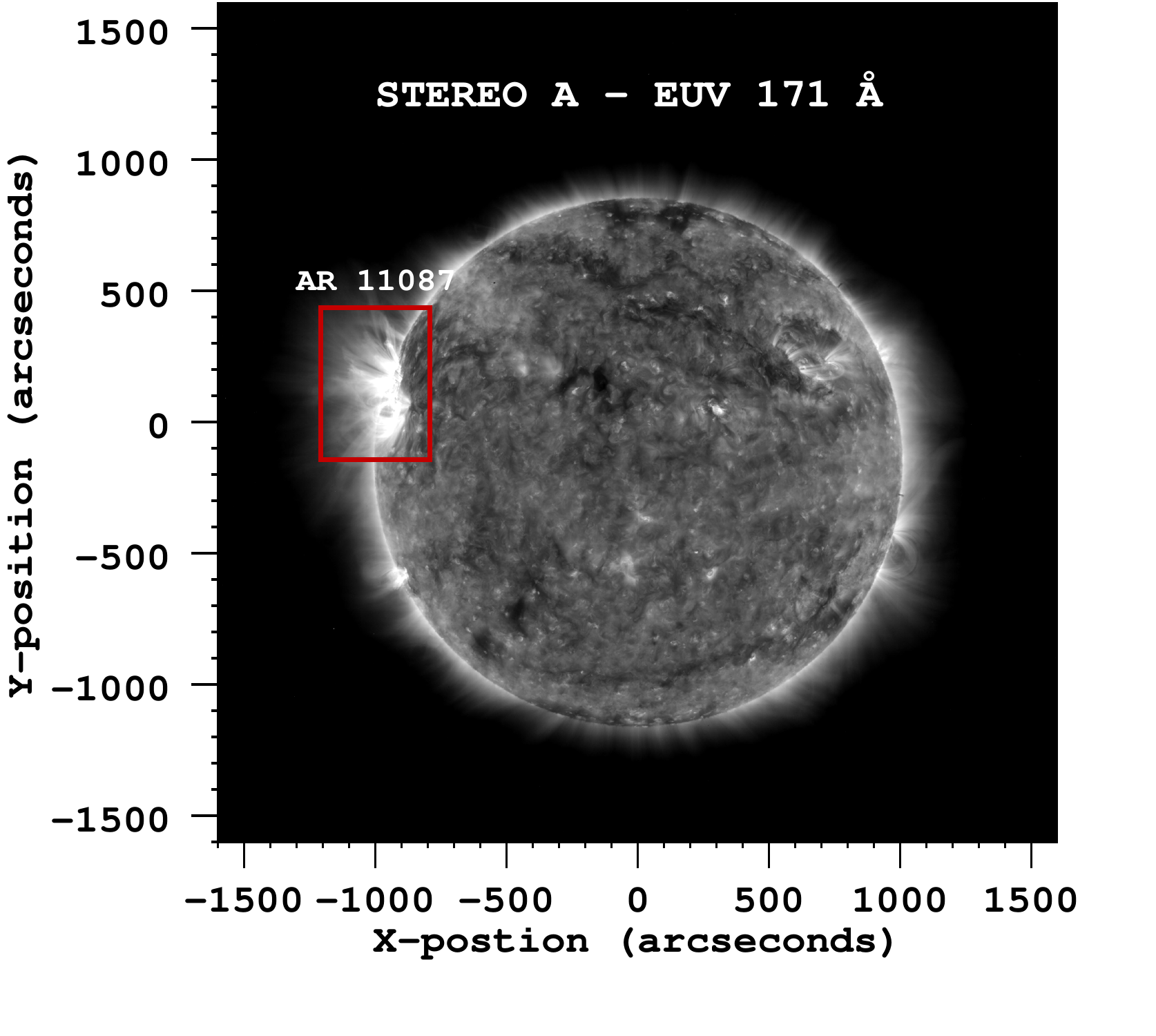}
		       }
		        \vspace{0.01\textwidth} 
 \caption{Images of the Sun with the active region AR 11087 from three different views observed on 2010 July 15 at 08:14 UT in 171 $\text{\AA}$  wavelength. The red rectangle marks the active region. In the left panel we display the EUVI/STEREO B image, in the middle panel, the AIA/SDO image and in the right panel, the EUVI/STEREO A image.}
\label{sc_pos}
\end{figure}

One of the criteria for selecting the data set was the separation angle between the two STEREO spacecraft.The stereoscopic reconstruction requires a separation angle between the view points larger than zero degrees and less than 180$^\circ$. For the selected event, the separation angle with respect to the center of the Sun between the two STEREO spacecraft was approximately 147$^\circ$, between STEREO A and SDO 77$^\circ$, between STEREO B and SDO 70$^\circ$ (Fig. \ref{sc_pos}) .
%%%%%%%%%%%%%%%%%%%%%%%%%%%%%%%%%%%%%%%%%%%%%%%%%%%%%%%%%%%%%%%%%%%%  
%%%%%%%%%%%%%%%%%%%%%%%%%%%%%%%%%%%%%%%%%%%%%%%%%%%%%%%%%%%%%%%%%%%%%%

%%%%%%%%%%%%%%%%%%%%%%%%%%%%%%%%%%%%%%%%%%%%%%%%%%%%%%%%%%%%%%%%%%%%     
\begin{figure}[h]
		\centerline{\hspace*{0.001\textwidth}
		    \includegraphics[width=0.7\textwidth, trim = 20 25 10 20, clip]{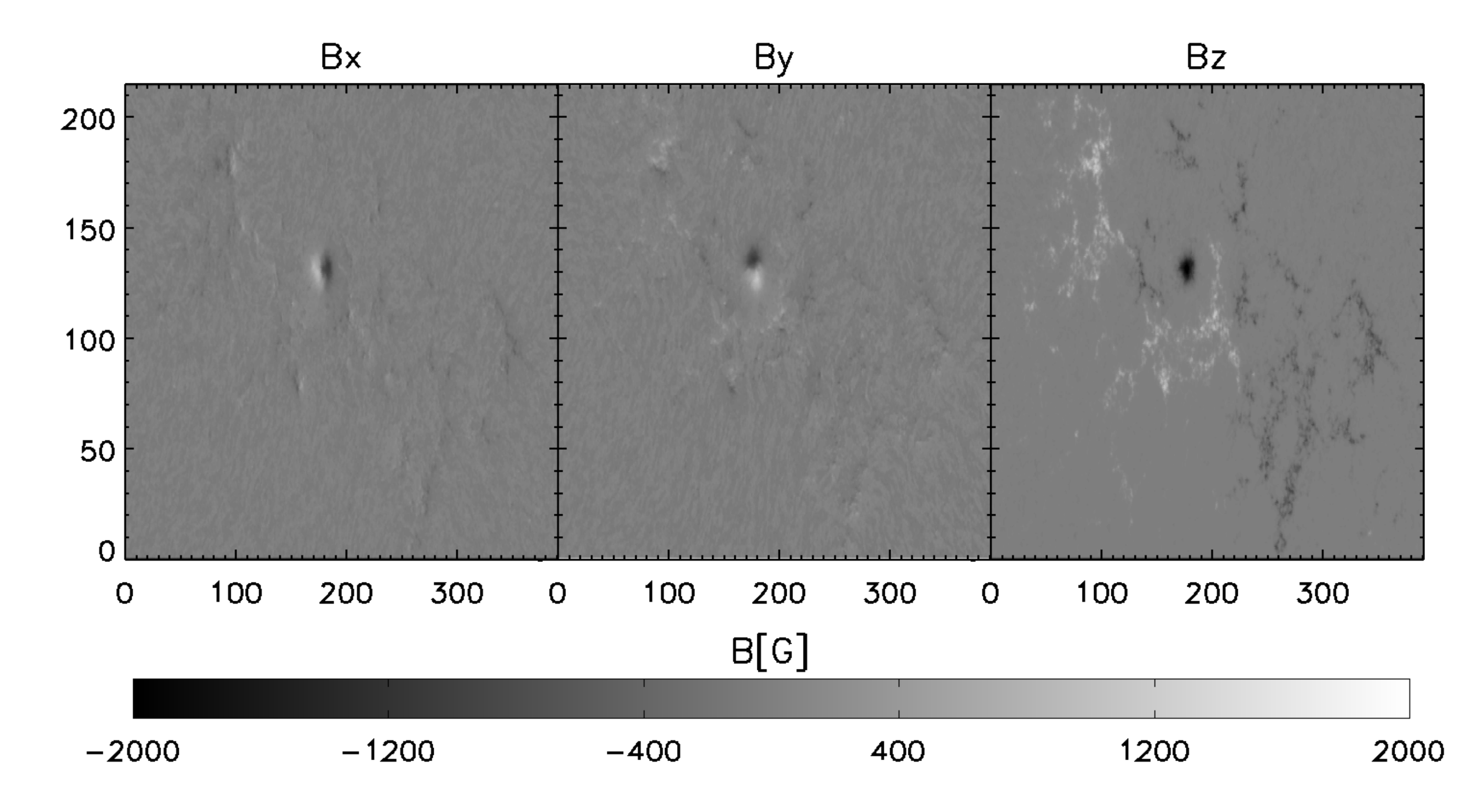}%trim=left bottom right top,clip
			    }   
 \vspace{0.01\textwidth} 
\caption{HMI/SDO vector magnetogram observed on 2010 July 15 at 08:14 UT.}
\label{mag_field}
\end{figure}
%%%%%%%%%%%%%%%%%%%%%%%%%%%%%%%%%%%%%%%%%%%%%%%%%%%%%%%%%%%%%%%%%%%%%%

Another selection criteria was the position of the active region on the solar surface as seen from the SDO spacecraft. As the accuracy of the photospheric field measurements become
strongly reduced towards the limb, we choose ARs close
to the disk center as seen from SDO (Fig. \ref{sc_pos}, middle panel).
A data set which fulfills these criteria is the active region AR 11087 observed on 2010 July 15.
We performed the 3D stereoscopic reconstruction using simultaneously extreme ultra-violet ($\lambda$ = 171 \AA{}) images recorded by the EUVI telescope onboard STEREO A and B and by the AIA telescope onboard SDO. The EUVI telescope has a FOV up to 1.7 R$_\odot$  ($\backsimeq$ 1182.7 Mm) and a spatial sampling of 1.6 arcsec pixel$^{-1}$ \citep{WuelserEtal2004}. AIA onboard SDO takes EUV images with a FOV of 1.5  R$_\odot$ and 0.6 arcsec pixel$^{-1}$ spatial sampling at each 12 seconds \citep{LemenEtal12}.
For the extrapolation of the NLFFF we used vector magnetograms provided by HMI/SDO (Fig. \ref{mag_field}).

\section{Data reconstruction}\label{sec:DataRecMBSR}
\subsection{Two and three view stereoscopic reconstruction}

One of the very important steps in 3D stereoscopic reconstruction is the correct identification and matching of the objects for reconstruction (e.g. coronal loops). In an ideal case, the objects for reconstruction have to be clearly visible and therefore easily identifiable. 

In many of the solar EUV observations the objects for reconstruction are not traceable in a straight forward manner. According to \cite{StenborgEtal2008} the major reasons for poor visualization of the data are the low contrast between the coronal structures and the background and the multiscale nature of the coronal features. Another reason is that in the EUV images we see the line-of-sight (LOS) integration of the radiation emitted by all the loops in a particular wavelength band. A variety of data processing procedures exists to enhance the visibility of the loop structures \citep{StenborgEtal2008}. The best method for our data processing we found to be the noise adaptive fuzzy equalization (NAFE) method developed by \cite{Druckmuller2013}. The method is based on histogram equalization and unsharp masking. We have applied this method for all of the three EUV images used in our 3D reconstructions.
%%%%%%%%%%%%%%%%%%%%%%%%%%%%%%%%%%%%%%%%%%%%%%%%%%%%%%%%%%%%%%%%%%%%     
\begin{figure}[h]
 \centerline{\hspace*{0.001\textwidth}
		     \includegraphics[ width=0.25\textwidth, trim = 20 20 20 30, clip]{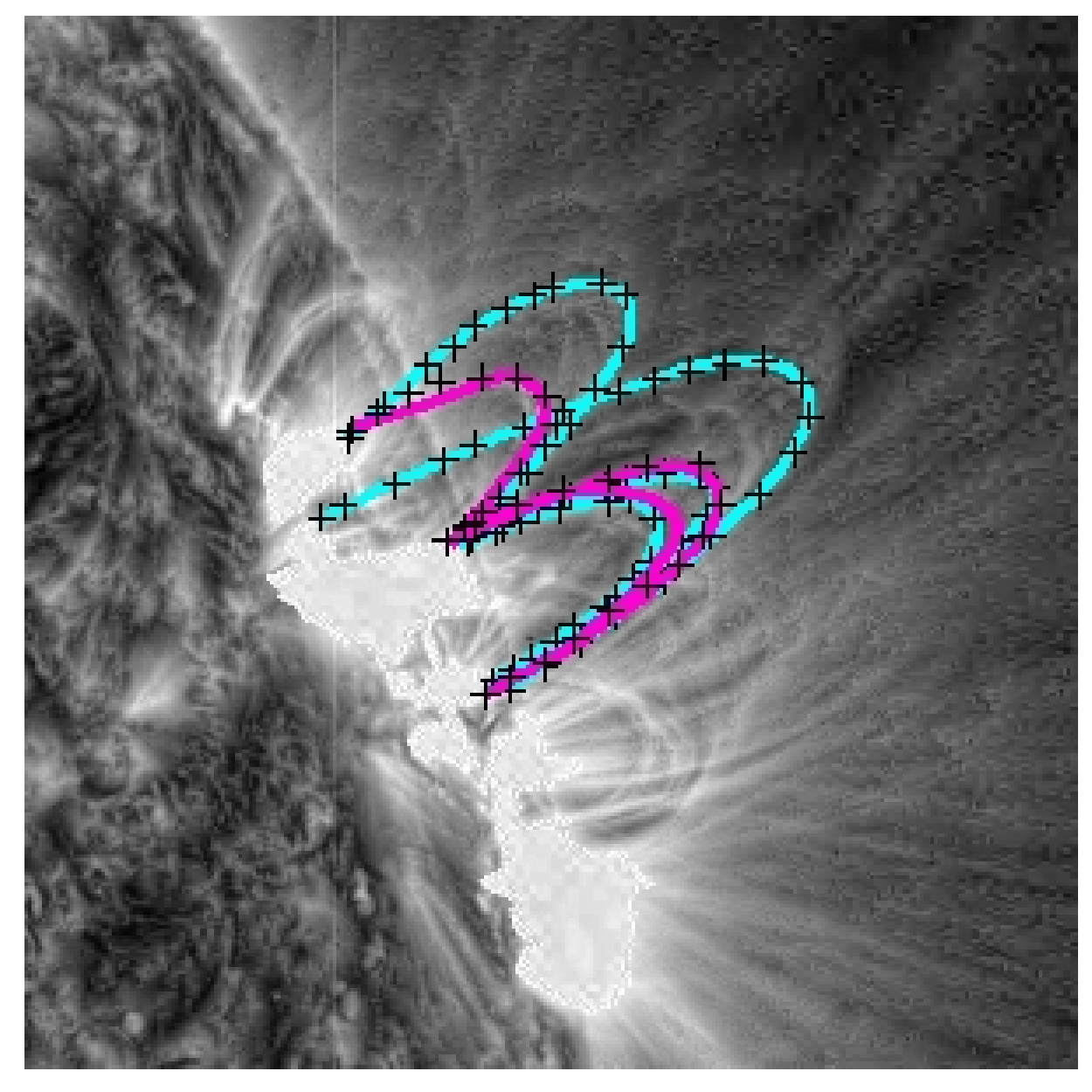} %trim=left bottom right top,clip
		       \includegraphics[width=0.41\textwidth, trim = 20 20 20 20, clip]{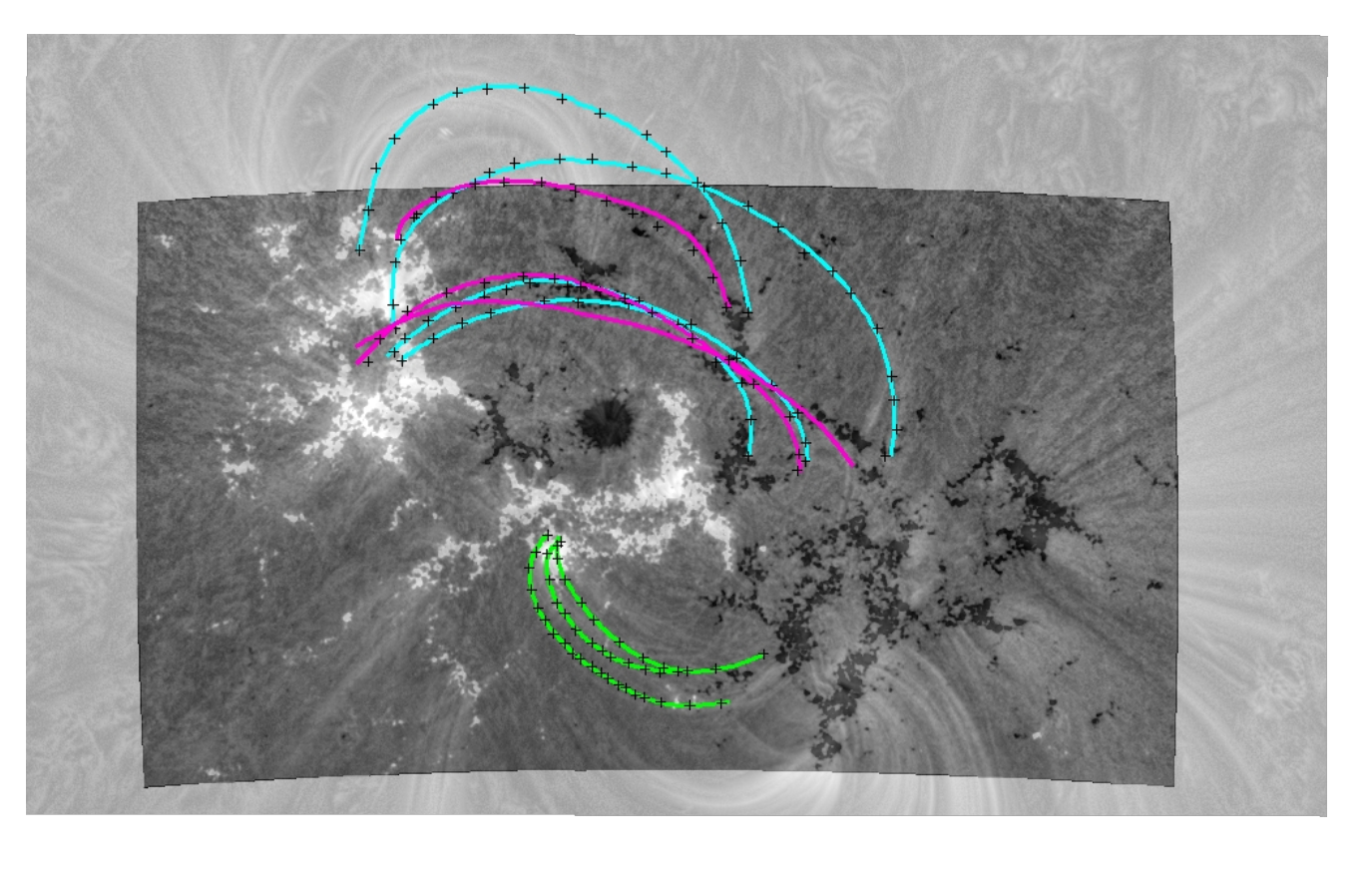}
		      \includegraphics[width=0.25\textwidth, trim = 20 20 20 30, clip]{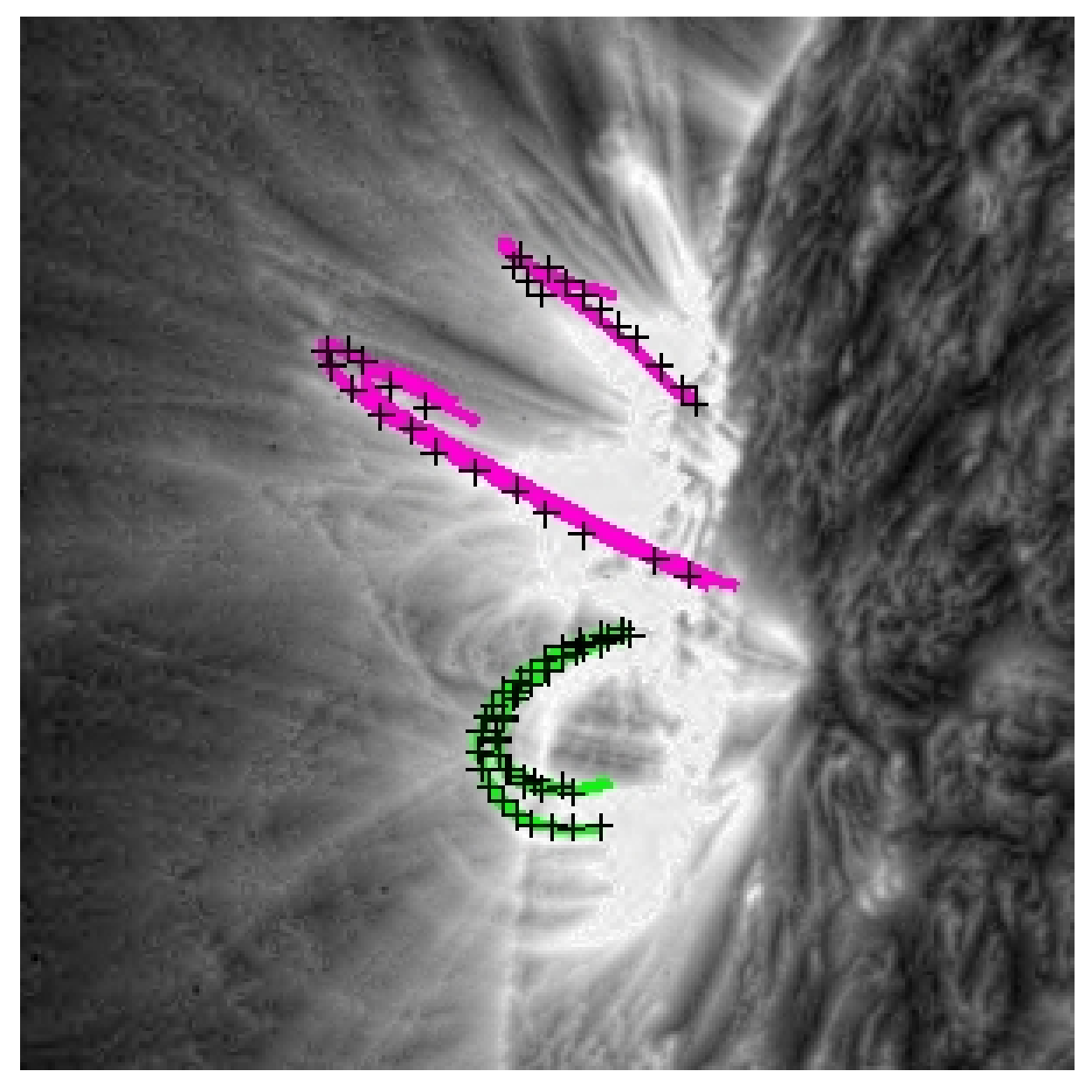}
		       }
		        \vspace{0.01\textwidth} 
 \caption{Projection of the 3D stereoscopically reconstructed loops overploted over the STEREO B (left panel), SDO (middle panel) and STEREO A (right panel). The magenta loops are reconstructed using all of the three spacecraft, the green loops are reconstructed using STEREO A and SDO and the light blue loops are reconstructed using STEREO B and SDO.}
\label{overplot_loops}
\end{figure}
%%%%%%%%%%%%%%%%%%%%%%%%%%%%%%%%%%%%%%%%%%%%%%%%%%%%%%%%%%%%%%%%%%%%%%

While some of the visualization problems can be resolved with image processing techniques, other problems such as saturated pixels cannot be resolved. In the data from STEREO A and B patches of saturated pixels restrained our identification and matching possibilities required by the reconstruction.
  
The configuration of the three spacecraft does not provide images with a visibility of the entire AR from all three vantage points simultaneously. Even though the data captured by the spacecraft fulfills our criteria of selection, the position of the three telescopes limits the number of loops which we can identify, trace and reconstruct. While the SDO satellite (see Fig. \ref{sc_pos}, middle panel) has a full view of the AR, the STEREO A (see Fig. \ref{sc_pos}, right panel) and B  (see Fig. \ref{sc_pos}, left panel) spacecraft were viewing a limited common area. In spite of all these above difficulties we could identify ten loops. Three loops were traced in all of the three images, three more loops in STEREO A and SDO and four loops in STEREO B and SDO. 

\begin{figure}[h]
		\centerline{\hspace*{0.001\textwidth}
		   \includegraphics[width=0.4\textwidth, trim = 0 0 0 15, clip]{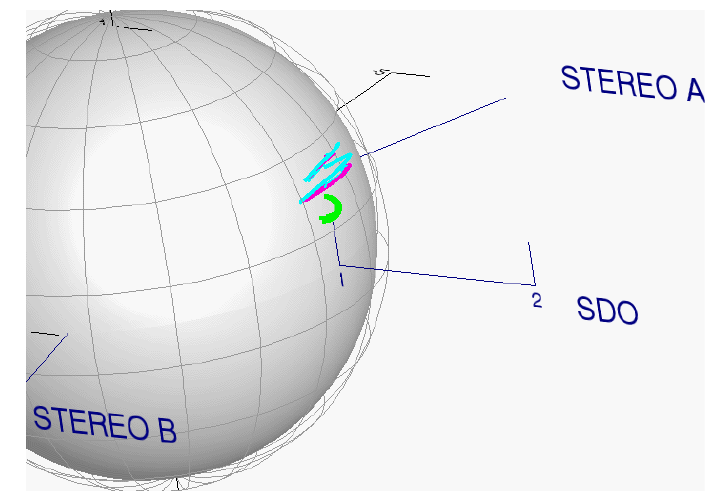}
           %trim=left bottom right top,clip
			    }   
 \vspace{0.01\textwidth} 
\caption{Solar toy model with the 3D reconstructed loops on top. The blue segments represents the direction towards the three spacecraft.}
\label{loops3D}
\end{figure}

In Fig. \ref{overplot_loops} we show the projection of the 3D stereoscopically reconstructed loops together with their tie-points (the black crosses) on each of the EUV images.
In Fig. \ref{loops3D} we present the 3D configuration of the Sun, represented as a gray sphere, and the direction of the three spacecraft together with the 3D reconstructed loops. The red loops are reconstructed using simultaneously all three spacecraft, the blue loops are reconstructed using the data from STEREO A and SDO while the green loops are based on the data from STEREO B and SDO. 
%%%%%%%%%%%%%%%%%%%%%%%%%%%%%%%%%%%%%%%%%%%%%%%%%%%%%%%%%%%%%%%%%%%%     
% \begin{tikzpicture}
% \end{tikzpicture}

%%%%%%%%%%%%%%%%%%%%%%%%%%%%%%%%%%%%%%%%%%%%%%%%%%%%%%%%%%%%%%%%%%%%%%

\subsection{S-NLFFF reconstruction}
\label{subsec:SNLFFF_rec}
The S-NLFFF reconstruction uses as input the photospheric vector-magnetograms provided by SDO/HMI and the 3D reconstructed loops described above. The HMI vector-magnetograms are mapped from the Helioprojective Cartesian to the Carrington Heliographic - Cylindrical Equal Area (CRLT/CRLN-CEA) coordinate system \citep{Bobra2014} in which we compute the 3D field reconstruction. The stereoscopically reconstructed loops were first calculated in HEEQ (Heliospheric Earth EQuatorial) coordinates and then mapped to the Carrington Heliographic coordinate system. 
%%%%%%%%%%%%%%%%%%%%%%%%%%%%%%%%%%%%%%%%%%%%%%%%%%%%%%%%%%%%%%%%%%%%     
\begin{figure}[h]
		\centerline{\hspace*{0.001\textwidth}
		    \includegraphics[width=0.5\textwidth, trim = 20 10 30 18, clip]{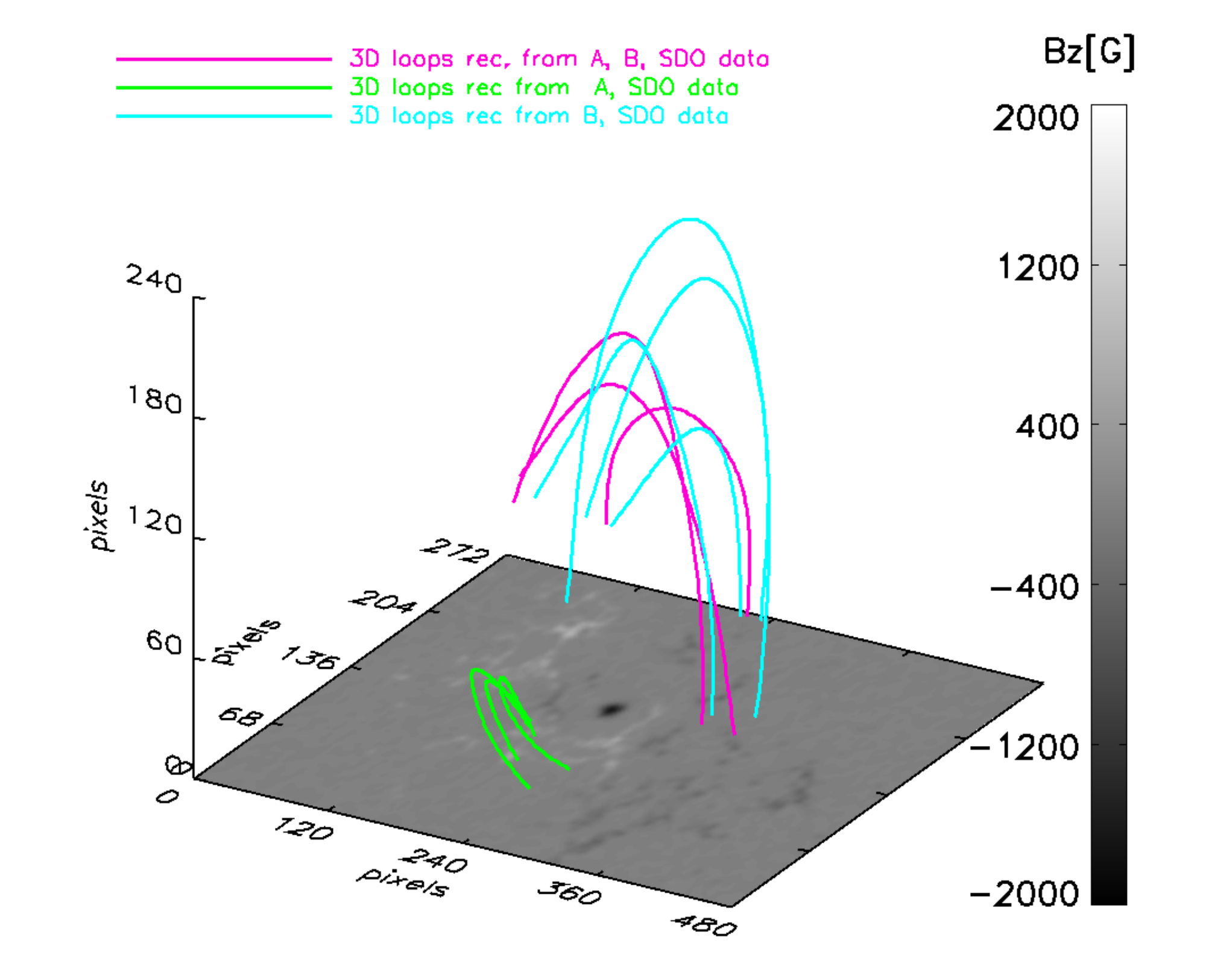} %trim=left bottom right top,clip
			    }   
 \vspace{0.01\textwidth} 
\caption{Plot of the 3D stereoscopically reconstructed loops inside the S-NLFFF computation box. At the bottom of the box, the radial component of the magnetic field is displayed.}
\label{vecmag_loops}
\end{figure}
%%%%%%%%%%%%%%%%%%%%%%%%%%%%%%%%%%%%%%%%%%%%%%%%%%%%%%%%%%%%%%%%%%%%%%

The computational box is 480$\times$272$\times$240 (pixels)$^3$ which is the equivalent of 350$\times$198$\times$175 (Mm)$^3$.
In the Fig. \ref{vecmag_loops} we show a 3D plot of the radial component of the magnetic field, color-coded at the bottom surface, along with the 3D stereoscopically reconstructed loops above.

The NLFF field reconstructions are calculated iteratively from an initial magnetic field until the field has relaxed to a force-free state. In order to find out how the final solution depends on the initial field and also to determine the impact of the loop data, we present alternative solution strategies.

Typically, the initial field for the iteration is the potential field  $\mathbf{B_\text{pot}}$ determined in the entire box from the normal component of the surface field. As an alternative, we iterate $\mathbf{B_\text{pot}}$ first on a coarse 240$\times$136$\times$120 grid and map the force-free field thus obtained from the coarse to the final 480$\times$272$\times$240 grid (so called multiscale approach). This interpolated force-free field is then used as initial field for the final iteration. For the coarse grid iteration, the boundary data is resampled accordingly from the original vector-magnetogram data. To see the effect of the loop data, we switch the loop constraint on, at different stages of the iteration.

We present here the result from five different setups
\begin{description}
\item[Setup 1] Starting from $\mathbf{B_\text{pot}}$ we iterate the force-free solution using the NLFFF on the final 480$\times$272$\times$240 grid without loop data. This is the conventional approach.
\item[Setup 2] Starting from $\mathbf{B_\text{pot}}$ we use S-NLFFF on the final grid, i.e., we include the loop data from the beginning of the iterations.
\item[Setup 3] We use the solution from Setup 1 as initial field for an iteration with S-NLFFF.
\item[Setup 4] We start from $\mathbf{B_\text{pot}}$ on the coarse grid and interpolate the coarse-grid force-free solution as initial field \textbf{ ($\mathbf{B^{coarse}_\text{NLFFF}}$)} for NLFFF on the final grid. No loop data is used.
\item[Setup 5] We use the interpolated coarse-grid field from Setup 4 as initial field \textbf{($\mathbf{B^{coarse}_\text{NLFFF}}$)} for S-NLFFF.
\end{description}

The natural approach would be to apply the S-NLFFF method on the fine grid (the Setup 2) and to evaluate the L$_\textit{1}$..L$_\textit{4}$ (Eq. \ref{L1}..\ref{L4}) and the angles between the magnetic field and the tangents of the 3D loops. 
We apply the S-NLFFF method to the Setup 2, 3 and 5 to see which one provides the best solution.
We run the Setup 3 to see if the force-freeness is maintained and  
in the same time the angles are minimized. 
\cite{2008SoPh..247..269M} claimed that the solution of the multiscale version of the NLFFF converges to a lower L (Eq. \ref{Ltot}) value when compared with the single grid solution. 
For this reason we considered the multiscale approach for the NLFFF and S-NLFFF method.

\section{Results}\label{sec:results}
\begin{figure}[h]

 		\centerline{\hspace*{0.001\textwidth}
		    \includegraphics[width=0.5\textwidth, trim = 20 220 60 270, clip]{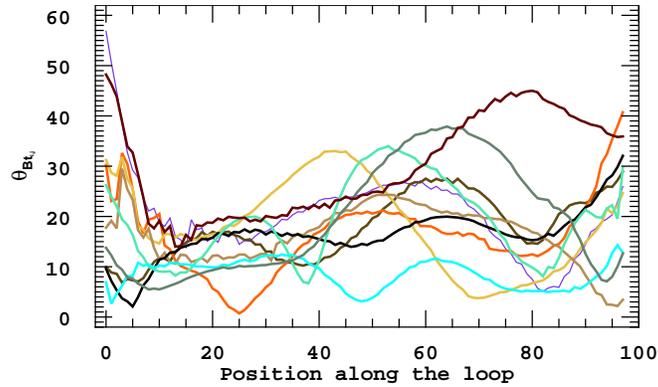}
			    }   
 \vspace{0.01\textwidth} 
 \caption{Angles between the NLFF magnetic field and the tangent of the 3D loops obtained as a result of Setup 1. Different colors represent different loops.}
\label{anglesNLFFF}
\end{figure}

\begin{figure}[h]
 \centerline{\hspace*{0.001\textwidth}
		     \includegraphics[ width=0.5\textwidth, trim = 18 220 60 265, clip]{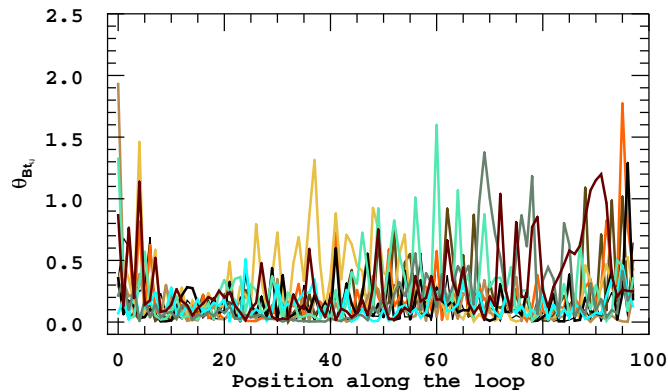} %trim=left bottom right top,clip
		  		       }
		        \vspace{0.01\textwidth} 
 \caption{The final angles between the S-NLFFF extrapolated magnetic field and the tangents of the 3D loops obtained as a result of Setup 5. Different colors represent different loops.}
\label{anglesSNLFFF}
\end{figure}

We calculated the angles ($\theta_{\textbf{Bt}_\textit{i,j}}$) between the magnetic field ($\mathbf{B}_\text{NLFFF}$) obtained with the NLFFF optimization method and the tangents ($\mathbf{t_\textit{i,j}}$, j=1...10, i=1..100) of the 3D stereoscopically reconstructed loops (see Fig. \ref{anglesNLFFF}, \ref{anglesSNLFFF}). The angles are calculated for each position i along the j$^{\text{th}}$ loop. Different colors represent different loops. The misalignment angles between $\mathbf{B}_\text{NLFFF}$ and $\mathbf{t_\textit{i,j}}$ are on average 20$^\circ$  and reach a maximum of approximately 60$^\circ$ (see Fig. \ref{anglesNLFFF}). The angles from Fig. \ref{anglesNLFFF} are obtained using $\mathbf{B}_\text{NLFFF}$ as a result of Setup 1, but the same profile is obtained using $\mathbf{B}_\text{NLFFF}$ from Setup 4.
%%%%%%%%%%%%%%%%%%%%%%%%%%%%%%%%%%%%%%%%%%%%%%%%%%%%%%%%%%%%%%%%%%%%     
%%%%%%%%%%%%%%%%%%%%%%%%%%%%%%%%%%%%%%%%%%%%%%%%%%%%%%%%%%%%%%%%%%%%     

By applying the S-NLFFF method, the angles $\theta_\mathbf{Bt_\textit{i,j}}$ between $\mathbf{B}_\text{S-NLFFF}$ and $\mathbf{t}_\textit{i,j}$ were reduced by a factor of more then 20 as shown in Fig. \ref{anglesSNLFFF}. For the calculation of the final angles $\theta_\mathbf{Bt_\textit{i,j}}$ from Fig. \ref{anglesSNLFFF} we used $\mathbf{B}_\text{S-NLFFF}$ as a result of Setup 5. Nevertheless, Fig. \ref{anglesSNLFFF} is representative also for the angles between the 3D loop tangents and the $\mathbf{B}_\text{S-NLFFF}$ obtained as a result of Setup 2 and 3.

%%%%%%%%%%%%%%%%%%%%%%%%%%%%%%%%%%%%%%%%%%%%%%%%%%%%%%%%%%%%%%%%%%%%   
%%%%%%%%%%%%%%%%%%%%%%%%%%%%%%%%%%%%%%%%%%%%%%%%%%%%%%%%%%%%%%%%%%%%%%

With the S-NLFFF method we could recover a magnetic field which is closer to the force-free condition. In Table \ref{table1} we present the values for the terms of the functional (see the detailed description of the terms in \cite{Wiegelmann04,Wiegelmann10,ChifuEtal2015}), namely the force-free ($L_\textit{1}$) term, the divergence of the magnetic field ($L_\textit{2}$) term, the closeness with the bottom boundary observation ($L_\textit{3}$) term and the closeness with coronal observable ($L_\textit{4}$) term. The residual values of the functional terms when applying the S-NLFFF method are lower than those obtained with the NLFFF method for the Setup 2 and 3 but a slightly larger for the Setup 5.

%%%%%%%%%%%%%%%%%%%%%%%%%%%%%%%%%%%%%%%%%%%%%%%%%%%%%%%%%%%%%%%%%%%%%%
\begin{table} [h]
\caption{The residual values of each of the functional terms 
%for each of the initial set up of the runs.
}
\label{table1}
 \centering
\begin{tabular}{c c c c c c c c }
\hline 
Configuration& No. grids & Initialization&Methods & $   L_1  $ & $  L_2   $ & $  L_3   $ & $  L_4  $ \\ \hline
Setup 1 & one & B$_\text{pot}$& NLFFF  & $  5.2  $ & $  3.2  $ & $  12.9 $ & $   -   $   \\ 
Setup 2 & one & B$_\text{pot}$ & S-NLFFF & $  4.6  $ & $  2.7  $ & $  12.2 $ & $ 0.0011$   \\
Setup 3 & one & B$_\text{NLFFF}$ &S-NLFFF & $  4.9  $ & $  3.0  $ & $  11.5 $ & $ 0.0041$   \\ 
Setup 4 & two & B$_\text{pot}$ &  NLFFF  & $  3.7  $ & $  2.2  $ & $  12.2 $ & $   -   $   \\
Setup 5 & two & B$^{coarse}_\text{NLFFF}$ & S-NLFFF & $  4.0  $ & $  2.3  $ & $  11.9 $ & $ 0.0007$   \\ 
\hline
\end{tabular}
\end{table}
%%%%%%%%%%%%%%%%%%%%%%%%%%%%%%%%%%%%%%%%%%%%%%%%%%%%%%%%%%%%%%%%%%%%%%

\begin{figure}[h]
 \centerline{\hspace*{0.001\textwidth}
		     \includegraphics[ width=0.6\textwidth, trim = 20 230 20 280, clip]{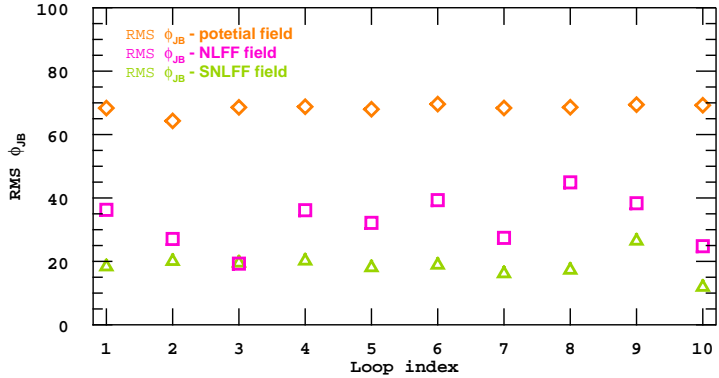} %trim=left bottom right top,clip
		  		       }
		        \vspace{0.01\textwidth} 
 \caption{Root mean square of the angles between the current and the potential field (orange rhombus), the extrapolated NLFFF (magenta squares) and the extrapolated S-NLFFF (green triangles) for each of the 3D loops.}
\label{anglesJB}
\end{figure}

We evaluated the angles ($\phi_\mathbf{JB}$) between the magnetic field and the current for each loop, along the loop. We derived the $\phi_\mathbf{JB}$ angles between the potential, NLFF and S-NLFF field and the current. For comparing the three cases, we calculated the root mean square (RMS) of the angles $\phi_\mathbf{JB}$ for each loop. This is a critical test because the current $\mathbf{J}$ is derived by differentiation from the magnetic field $\mathbf{B}$ which amplifies the noise, especially where the field strength is low. In Fig. \ref{anglesJB} we show the RMS of $\phi_\mathbf{JB}$ for each loop. Here we present the angles derived using the $\mathbf{B}_\text{NLFFF}$ obtained as a solution of Setup 1 and the $\mathbf{B}_\text{S-NLFFF}$ obtained as a solution of Setup 5. The evolution of the $\phi_\mathbf{JB}$ from Fig. \ref{anglesJB} is representative also for angles derived using the NLFFF solution of Setup 4 and the S-NLFFF solution of setups 2, 3 and 5. The current is more aligned with the magnetic field after using the reconstructed 3D loops as constrain for the S-NLFFF method. 

%%%%%%%%%%%%%%%%%%%%%%%%%%%%%%%%%%%%%%%%%%%%%%%%%%%%%%%%%%%%%%%%%%%%     
%%%%%%%%%%%%%%%%%%%%%%%%%%%%%%%%%%%%%%%%%%%%%%%%%%%%%%%%%%%%%%%%%%%%
\section{Discussions}\label{sec:dicussions}

\cite{2009ApJ...696.1780D} compared different coronal NLFFF models with EUV coronal loops observations. The conclusion of the study was that the misalignment angles between the extrapolated NLFF field and the 3D stereoscopically reconstructed loops reaches a maximum of approximately 45$^\circ$. In agreement with the results of \cite{2009ApJ...696.1780D} we derived similar angles between the magnetic field ($\mathbf{B}_\text{NLFFF}$) obtained with the NLFFF optimization method (for Setup 1 and 4) and the tangents $\mathbf{t}_{i,j}$ of the 3D stereoscopically reconstructed loops (see Fig. \ref{anglesNLFFF}).

In a previous paper \citep{ChifuEtal2015} we presented and tested the S-NLFFF method with semi-analytic data. The results of the tests predicts that the S-NLFFF method is capable of reducing the values of the $\theta_{\mathbf{Bt}_\textit{i,j}}$ angles below 2$^\circ$. In all of the cases studied in this paper, the S-NLFFF method was capable to reduce the angles even further (see Fig. \ref{anglesSNLFFF}).   

In an ideal case, the residual values of the functional terms L$_\textit{1}$..L$_\textit{4}$ (Eq. \ref{L1}...\ref{L4}) would be zero. Since the observational data contains errors and the magnetic field model is based on certain assumptions, the residual values cannot exactly be minimized to zero. The smaller the residual value L$_\textit{1}$ (Eq. \ref{L1}), the close is the field to the force-free condition. For the setups 2 and 3, the S-NLFFF could bring the magnetic field closer to a force-free solution when compared with the reference field (Setup 1). 

From the evaluation of the root mean square angles ($\phi_\mathbf{JB}$) between the current and the magnetic field, we could see an improvement in the average alignment for all of the three setups. The large values in the angle between the force-free magnetic field and current are probably due to the large uncertainties in the horizontal vector field component, in particular in the weak regions of magnetic field. Even for the Setup 5 for which the residual values for the force-free terms did not improve when applying S-NLFFF, the average angle along the loop between the field and the current became smaller. Over all we can say that the new method which includes the constraints from the corona improves not only the agreement between modeling and observations, but it also improves the force-freenes of the obtained magnetic field.

For most of the 3D stereoscopically reconstructed loops used as constraint for the magnetic field, the S-NLFFF method is able to reduce the angles between the magnetic field and the 3D loop tangents below 2$^{\circ}$. Nevertheless, there are few loops for which the angles between $\mathbf{B}_\text{S-NLFFF}$ and $\mathbf{t}_i$ remain large after S-NLFFF treatment. These loops have a deviation of $\gtrsim$ 65$^{\circ}$ when compared with the NLFFF model field (Setup 1 and 4). When this field was used as initial condition for S-NLFFF (Setup 3) the average angle could be reduced by a factor of 2-10 but not below 5$^{\circ}$.

In this paper we present the performance of the S-NLFFF method using ten 3D coronal loops as a constraint for modeling the coronal magnetic field. For these ten loops we show that the S-NLFFF method can obtain a good agreement between the modeled coronal magnetic field and the coronal loops observations. The S-NLFFF method can also obtain a much better alignment between the current and the magnetic field which is an indication that we obtain a better field in terms of force-freenes. The residual value of force-free integral value (Eq. \ref{L1}) decreases only little. The reason is probably that the few loops we included improve the field in their local environment but have limited impact on metrics which average over a much larger volume. We believe that more loops which occupy a larger fraction of the computational box will also improve the quality measures over the entire box.would have a larger impact on the improvement of the force-freenes.
%For their extrapolation they used a number of 26 loops. The method was also applied before and after a flare \citep{2014ApJ...783..102M}. For this application they used around 150 loops. }
\vspace{10pt}

Data are courtesy of NASA/SDO and the AIA and HMI science teams. The authors thank the STEREO SECCHI consortia for supplying their data. STEREO is a project of NASA. I.C is grateful to Hans-Peter Doerr for helpful suggestions. This work was supported by DLR fund 50 OC 1301

%\bibliography{apj1.bib}
\bibliography{apj1.bib,bibtw.bib}
\bibliographystyle{aasjournal}
\end{document}